\documentstyle[twoside,fleqn,espcrc2,epsf]{article}

\newcommand{\be}{\begin{equation}}
\newcommand{\ee}{\end{equation}}
\newcommand{\vap}{\varepsilon}

\title{New Analytical Results on Anisotropic Membranes}
\author{Mark Bowick ${}^{\rm a}$\thanks{\tt bowick@physics.syr.edu} 
and {\underline {Alex Travesset}} \thanks{\tt alex@suhep.phy.syr.edu}
\address{ Physics department, Syracuse University, \\
Syracuse, NY 13244-1130, USA \\ }
}
       
\begin{document}

\begin{abstract}
 
We report on recent progress in understanding the tubular phase of
self-avoiding anisotropic membranes. After an introduction to the problem,
we sketch the renormalization group arguments and symmetry considerations
that lead us to the most plausible fixed point structure of
the model. We then employ an $\vap$-expansion about the
upper critical dimension to extrapolate to the physical
interesting 3-dimensional case. The results are $\nu=0.62$ for the
Flory exponent and $\zeta=0.80$ for the roughness exponent. Finally we
comment on the importance that numerical tests may have to test these 
predictions.

\end{abstract}

\maketitle

\section{Introduction}

The statistical properties of D-dimensional objects embedded
in $d$-dimensional space have been the subject of intense 
analytical and numerical work in the last ten years.
An introduction to the problem, as well as an update with some
recent results has been already presented in M. Bowick's
talk \cite{MJB} and in the plenary talk \cite{GTH}. These studies
are of direct experimental interest for cases such as
$(D=2,d=3)$ (membranes) or $(D=1,d=3)$ (polymers) 
(see \cite{MJB}).

In \cite{RT1} it was shown that anisotropy has a remarkable effect in
a model of phantom crystalline surfaces; there is a new phase, the tubular
phase, characterized by being flat in one internal direction
and crumpled in the other ones. This new phase has 
been nicely corroborated by numerical simulations \cite{BFT}, see
\cite{MJB} for an update.

\begin{figure}
\epsfxsize=1.8in \centerline{\epsfbox{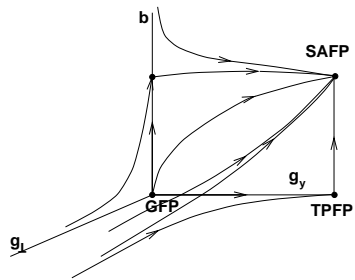}}
\caption{The phase diagram for self-avoiding anisotropic membranes
with the Gaussian fixed Point (GFP), the phantom tubular fixed point
(TPFP) and the self-avoidance fixed point (SAFP).}
\label{fig_BG}
\end{figure}

\begin{figure}
\epsfxsize=1.8in \centerline{\epsfbox{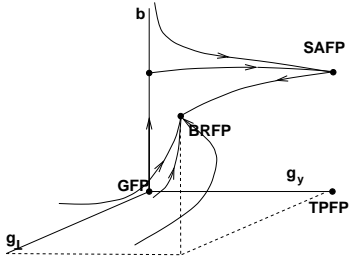}}
\caption{The phase diagram for self-avoiding anisotropic membranes
with the Gaussian fixed Point (GFP), the phantom tubular fixed point
(TPFP), the self-avoidance fixed point (SAFP) and the bending rigidity
fixed point (BRFP).}
\label{fig_RT}
\end{figure}

While the phantom membrane model is completely 
understood, both analytically and numerically, the situation for the more
physical self-avoiding case has been rather controversial. Once
the self-avoidance perturbation is added to the phantom
model it was found in \cite{BG} that the large distance properties 
of self-avoiding anisotropic
membranes are described by a new Fixed Point (the SAFP), different from
the phantom one (the TPFP), but 
perturbative in $\vap=11-d$. The phase diagram implied is shown 
in fig.~\ref{fig_BG}. In \cite{RT2} it was argued that the SAFP 
is infrared unstable, and consequently, the
large distance properties of anisotropic self-avoiding membranes 
were described by a new Fixed Point (BRFP), which has a non-trivial
anomalous dimension for the bending rigidity term. The phase diagram
implied in \cite{RT2} is the one depicted in fig.~\ref{fig_RT}.

In this talk we report on new analytical results \cite{US}. The aim 
of this work is twofold; first to clarify the phase diagram
of the model, and second to compute the critical exponents which
provide predictions to be tested numerically or even in actual 
experiments.

\section{The Phase diagram}

The configuration of a membrane is described by giving the position
${\vec r}({\bf x})$, in the $d$-dimensional embedding space, of a point
in the membrane labeled by a $D$-dimensional internal coordinate 
${\bf x}$. In the tubular phase, where the membrane is crumpled
in $D-1$ dimensions (denoted by ${\bf x}_{\perp}$) and flat in
a distinguished direction $y$, $\vec r$ is expanded as
\be\label{MONG}
\vec r({\bf x})=(\zeta_y y+u({\bf x}),\vec h({\bf x})).
\ee
The relevant degrees of freedom are then in-plane phonons $u$, and
out-of-plane ones $\vec h$. Let us remark that those fields 
have different engineering dimensions, namely, different rescalings in 
the language of the Wilsonian renormalization group.

The most general free energy for this system was first written in \cite{RT1}.
It is constructed as a derivative expansion in $\vec r$ together with the
requirements of translational and rotational symmetry. A direct general 
analysis of this free Energy is a very difficult task, close to impossible
if self-avoidance is considered. Our interest, fortunately, is focused 
on the universal
properties of the model, that is, its large distance properties. The challenge
becomes to correctly identify the terms that do not affect the universal
properties (the irrelevant ones). There is a powerful tool available for
that task, the Renormalization Group. 

The implementation of the Renormalization Group in this particular problem 
presents new and interesting features. For example, the rotations of the 
tubules imply that the free energy should be invariant under
\be\label{exact_rot}
\begin{array}{l l} u \rightarrow  & u \cos \theta + \sin \theta h +
(\cos \theta -1)y \\
h \rightarrow   & h \cos \theta - \sin \theta u -\sin \theta y
\end{array} \ .
\ee
This symmetry mixes the $\vec h$ and $u$ fields, which, as already pointed out,
have different rescalings. It should not come as a surprise then, that
the large distance realization of the $O(D-1)$ rotational symmetry is
different from Eq.~\ref{exact_rot}, \cite{US}
\be\label{inv_gen_infty}
\begin{array}{l l} u \rightarrow  & u + {\vec A} {\vec h}
-\frac{1}{2}{\vec A}^2 y +{\cal O}(e^{2(\nu-z)l}) \\
{\vec h} \rightarrow   & {\vec h}-{\vec A} y+{\cal O}(e^{2(\nu-z)l})
\ , \\
\end{array}
\ee
where $\vec A$ is an arbitrary $D-1$ dimensional vector.

The symmetry Eq.~\ref{inv_gen_infty} provides an important guiding
principle to elucidate the phase diagram. A more detailed analysis
performed in \cite{US} shows that the most
plausible phase diagram when self-avoidance is included is the one in 
fig.~\ref{fig_BG}. Of course, more complicated situations (reminiscent of
the ones depicted in fig.~\ref{fig_RT} are not completely  ruled out, 
but we do not find enough evidence to sustain them. A further clarification 
for this debate definitely requires a full treatment of both nonlinear 
elastic terms and self-avoidance, an open problem so far. 

\section{Critical exponents}

The self-avoidance perturbation is relevant for any embedding 
dimension $d<d_c^{SA}$, where
\be\label{dim_self}
d_c^{SA}(D)=\frac{6D-1}{5-2D} \ .
\ee
The fixed point of interest, the SAFP, is perturbative in $\vap$, with
$\vap(D,d)=3D-\frac{1}{2}-(\frac{5}{2}-D)d$.

In \cite{BG} a direct $\vap$-expansion was performed at $D=2$. In that case,
$\vap(2,d)=\frac{11-d}{2}$ and the extrapolation to the physical interesting
case $d=3$ was not found to be robust against higher order
corrections (recall that $\vap(2,3)=4$). The values for the critical
quantities obtained in \cite{BG} show a very large uncertainty. On the 
other hand, we have emphasized how important is to get good estimates of 
the critical indices at the SAFP.

In \cite{US}, the one loop critical exponents are computed 
for arbitrary internal dimension $D$
( For those values of $D$ for which the model 
is well defined are $3/2< D <5/2$). This generalization provides 
a new extrapolation parameter, and allows for generalized  
$\vap$ expansions \cite{EXT} that may deliver reliable results
at $D=2$. 

The extrapolation techniques we used are rather sophisticated. Essentially
they consist in re-expressing the critical exponents in new variables
such that the next corrections get minimized. We tested different 
variables and different corrections to the exponents.
We refer the interested reader to the original work \cite{US} for
details. We just quote
the results for the two main critical exponents coming from the 
best extrapolations, which are
listed in Table~\ref{tab__EXP_fin}. The results
are compared with the uncontrolled Flory estimate, which is usually a
good approximation.

\begin{table}[htb]
\centerline{
\begin{tabular}{|c||l|l||l|l|}
\multicolumn{1}{c}{$d$}   &
\multicolumn{1}{c}{$\nu$}   & \multicolumn{1}{c}{$\nu_{Flory}$} &
\multicolumn{1}{c}{$\zeta$} & \multicolumn{1}{c}{$\zeta_{Flory}$} \\\hline
8  & $0.333(5) $ & $0.333$ & $0.60$  & $0.600$  \\\hline
7  & $0.374(8) $ & $0.375$ & $0.64$  & $0.643$  \\\hline
6  & $0.42(1)  $ & $0.429$ & $0.68$  & $0.692$  \\\hline
5  & $0.47(1)  $ & $0.500$ & $0.72$  & $0.750$  \\\hline
4  & $0.54(2)  $ & $0.600$ & $0.76$  & $0.818$  \\\hline
3  & $0.62(2)  $ & $0.750$ & $0.80$  & $0.900$  \\\hline
\end{tabular}}
\caption{Final results for critical exponents. $\nu$ is the Flory exponent
and $\zeta$ is the roughness exponent. The Flory estimate is 
quoted as $\nu_{Flory}$.}
\label{tab__EXP_fin}
\end{table}

\section{Conclusions and Outlook}

In this talk we discussed the phase diagram for anisotropic
membranes. We have definite analytical predictions for the critical 
exponents characterizing the tubular phase of self-avoiding membranes. 
At this stage, further refinement would necessitate a two-loop calculation 
for arbitrary $D$ that, as a byproduct, would provide a valuable check 
for our extrapolation.

There are other aspects which are worth pointing out. As discussed in 
\cite{MJB}, existing arguments favors the belief that the crumpled phase 
for self-avoiding isotropic membranes disappears whenever bending 
rigidity terms are present, although a definite proof is still
lacking. One might legitimately ask if the same is true for the tubular
phase. Intuitively, one would think that this is not the case, since
self-avoidance is less constraining in this case. In fact, the arguments
that lead to the considerations above do not apply( and the corresponding 
molecular dynamics simulations are absent). Our conclusion is that
the tubular phase should be observed both
in numerical simulations and actual experiments so that
our predictions can be tested.

There has been some progress in the numerical analysis of this model
as reported in \cite{MJB}. There is evidence that the tubular phase 
does survive, but the available data is not still enough to enable 
reliable estimates for critical exponents. Anyway, we hope that the
calculations presented here will inspire further work on the numerical 
side. Further progress in understanding the self-avoiding tubular phase 
very much requires the insight of numerical work.

The research of M.B. and A.T. has been supported by the U.S. 
Department of Energy under contract DE-FG02-85ER40237.

\end{document}